\documentclass[aip,rsi,preprint]{revtex4-1} 
\usepackage{siunitx}
\usepackage{graphicx}
\usepackage{epstopdf}
\usepackage[colorinlistoftodos,prependcaption,textsize=tiny]{todonotes}
\usepackage{amsmath}
\usepackage{booktabs}
\usepackage[caption=false]{subfig}

\let\originaleqref\eqref
\renewcommand{\eqref}{Equation~\originaleqref}

\newcommand{\figref}[1]{Figure~\ref{#1}}
\newcommand{\tabref}[1]{Table~\ref{#1}}
\newcommand{\etal}{\textit{et al.}}

\draft 

\begin{document}


\title{Meta-instrument: high speed positioning and tracking platform for near-field optical imaging microscopes } 



\affiliation{Department of Optomechatronics, Netherlands Organisation for Applied Scientific Research TNO, Stieltjesweg 1, 2628CK, Delft, The Netherlands}
\affiliation{Department of Optics, Netherlands Organisation for Applied Scientific Research TNO, Stieltjesweg 1, 2628CK, Delft, The Netherlands}
\affiliation{Structural Optimization and Mechanics, Faculty of Mechanical Engineering, Delft University of Technology, Mekelweg 2, 2628CD, Delft, The Netherlands}

\author{R.J.F. Bijster}
\affiliation{Department of Optomechatronics, Netherlands Organisation for Applied Scientific Research TNO, Stieltjesweg 1, 2628CK, Delft, The Netherlands}
\affiliation{Structural Optimization and Mechanics, Faculty of Mechanical Engineering, Delft University of Technology, Mekelweg 2, 2628CD, Delft, The Netherlands}


\author{R.W. Herfst}
\affiliation{Department of Optomechatronics, Netherlands Organisation for Applied Scientific Research TNO, Stieltjesweg 1, 2628CK, Delft, The Netherlands}

\author{J.P.F. Spierdijk}
\affiliation{Department of Optomechatronics, Netherlands Organisation for Applied Scientific Research TNO, Stieltjesweg 1, 2628CK, Delft, The Netherlands}

\author{A. Dekker}
\affiliation{Department of Optomechatronics, Netherlands Organisation for Applied Scientific Research TNO, Stieltjesweg 1, 2628CK, Delft, The Netherlands}


\author{W.A. Klop}
\affiliation{Department of Optomechatronics, Netherlands Organisation for Applied Scientific Research TNO, Stieltjesweg 1, 2628CK, Delft, The Netherlands}

\author{G.F.IJ. Kramer}
\affiliation{Department of Optomechatronics, Netherlands Organisation for Applied Scientific Research TNO, Stieltjesweg 1, 2628CK, Delft, The Netherlands}



\author{L.K. Cheng}
\affiliation{Department of Optics, Netherlands Organisation for Applied Scientific Research TNO, Stieltjesweg 1, 2628CK, Delft, The Netherlands}

\author{R.A.J. Hagen}
\affiliation{Department of Optics, Netherlands Organisation for Applied Scientific Research TNO, Stieltjesweg 1, 2628CK, Delft, The Netherlands}

\author{H. Sadeghian}
\email[]{hamed.sadeghianmarnani@tno.nl}
\affiliation{Department of Optomechatronics, Netherlands Organisation for Applied Scientific Research TNO, Stieltjesweg 1, 2628CK, Delft, The Netherlands}


\date{\today}

\begin{abstract}
High resolution and high throughput imaging are typically mutually exclusive. The meta-instrument pairs high resolution optical concepts such as nano-antennas, superoscillatory lenses and hyperlenses with a miniaturized opto-mechatronic platform for precise and high speed positioning of the optical elements at lens-to-sample separations that are measured in tens of nanometers. Such platform is a necessary development for bringing near-field optical imaging techniques to their industrial application. Towards this purpose, we present two designs and proof-of-principle instruments that are aimed at realizing sub-nanometer positional precision with a \SI{100}{\kHz} bandwidth. 
\end{abstract}

\maketitle 

\section{Introduction}
Resolution is the de-facto standard by which microscopes are compared. The common desire of detecting ever smaller features has heralded a long chain of developments and a search for better techniques and instrumentation; feats that are deemed important to the present day\cite{NobelMediaAB20142014}. 

Although scanning probe microscopes provide laboratories with atomic resolution - far exceeding the capabilities of contemporary (far-field) optical techniques - the desire to increase the performance of optical microscopes remains. These microscopes give direct access to a large field of view, which is a distinct advantage of direct capture techniques over scanning probe technologies. Moreover, applications involving life samples or damage prone features, e.g., high aspect ratio FinFETs\cite{Hisamoto2000}, require non contact microscopy technologies, despite the progress that is made in reducing the interaction forces in scanning probe techniques\cite{Keyvani2015a,Sadeghian2015b}.

In recent years solid immersion lenses\cite{Wu1999} (SIL), super-oscillatory lenses (SOL)\cite{Berry2009}, hyperlenses\cite{Lu2012a} and nano-antennas\cite{Novotny2011} have emerged as promising concepts for improved optical resolution. However, these concepts share more than just their sub-diffraction-limited performance. Hyperlenses, SILs and nano-antennas require very precise positioning above the sample at distances that are measured in tens of nanometres. For this reason these optical elements have largely been subject of theoretical study and preliminary experimental verification with \textit{fixed} geometries\cite{Fang2005, Melville2005, Melville2006, Lee2005, Liu2007, Liu2007b, Liu2007c, Huang2007a, Huang2007, Jeppesen2009, Chaturvedi2010, Rogers2012, Rogers2013, Rogers2013a, Roy2013, Wong2013a, Roy2014}. Bringing these optical techniques from the laboratory to their industrial applications requires that they are paired with a mechatronic platform, that is capable of positioning the lens 1) at the required distance from the sample, 2) with sub-nanometer precision and resolution and 3) at high scanning speed for high-throughput imaging. 

Towards developing exactly such a platform, we present in this manuscript the system architecture that we have dubbed the 'meta-instrument' and two designs of such an apparatus.

The paper is organized such to present the system architecture and requirements in Section~\ref{sec:architecture}, followed by a discussion of the current realization and the second generation in Sections~\ref{sec:current}~and~\ref{sec:future}, respectively.

\section{Architecture}
\label{sec:architecture}
\begin{figure}[t]
 	\centering
 	\includegraphics[width=6.5cm]{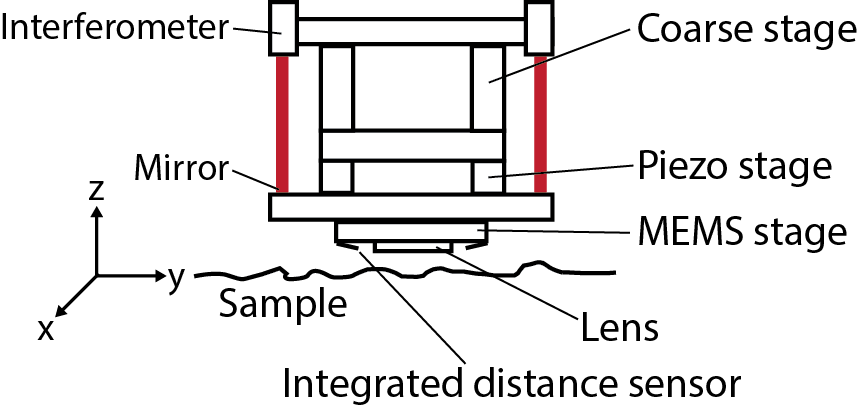}
 	\caption{\label{fig:metaimpression}A diagram representing the instrument in proximity to the sample. The z-direction is the approach direction, while the xy-plane is the scanning plane.}%
 \end{figure}

Before discussing the architecture itself, it is instructive to consider a typical use case for an imaging microscope. Operation starts with sample placement. Safe placement of a sample by an operator implies a need for sufficient clearance between the lens and the sample. This is achieved by placing the instrument in a stand-by position at millimeters from the sample surface. Once the sample is loaded and secured in place, the instrument has to engage the sample and position the imaging element at the correct working distance. To allow imaging of an area larger than the field of view of the lens, the sample and lens have to be scanned relative to each other. This configuration is depicted in \figref{fig:metaimpression}. During scanning, small height variations on the sample may cause the lens to be placed out-of-focus or  may even result in physical contact that causes damage to either lens or sample. To avoid such events, the distance to the sample is continuously measured and controlled. Once imaging is finished, the instrument has to retract from the surface to provide sufficient clearance for further handling of the sample.  

Based on this and comparable scenarios, it is clear that at a system level the instrument platform has to fulfill the functions of 1) engaging the optical element with the sample, 2) tracking the surface at constant height to compensate for variations in sample topography and 3) correct for any misalignment between lens and sample to avoid contact with the sample. To achieve this functionality the instrument is composed of several functional groups of distinct purposes. 

\subsection{System requirements}
\label{sec:systemreqts}
In terms of imaging and inspection needs, the semiconductor industry poses some of the most demanding requirements. The decreasing critical dimensions of transistor gates increasingly strains inspection and metrology capabilities: more devices need to be inspected more often during production to guarantee that properly working devices are created. The industry predicts that this will become a bottleneck for high volume manufacturing\cite{Bunday2016} if the resolution and throughput of inspection and metrology tools do not increase.

The requirements imposed on the instrumentation platform described in this manuscript are summarized in \tabref{tab:requirements}. The corresponding rationales are explained below.

The considered optical elements (SIL, SOL, hyperlens) provide a typical field-of-view in the order of $\SI{10}{\um}\times\SI{10}{\um}$. To be able to image areas that are larger than this field-of-view, a scanning mode of operation is necessary in which the lens moves parallel to the sample surface. Industrial parties have indicated that a scan speed of \SI{1}{\mm\per\s} to \SI{100}{\mm\per\s} is required for any optical tool to be relevant in a high volume manufacturing environment\cite{Bunday2016}. Because the lens-to-sample separation can be as small as \SI{10}{\nm}, surface contamination, surface roughness and local variations in the sample topography can result in dramatic changes in the lens-to-sample separation over small scan lengths. We assume that these variations can occur over distances of \SI{10}{\nm}. Combined with the required scan speed, this yields a required bandwidth $\ge\SI{100}{\kHz}$.

To guarantee the required optical performance, the lens-to-sample separation has to be kept within bounds. For the considered optical elements, a position precision of the piston motion of \SI{1}{\nm} is desired. However, a piston motion alone does not suffice to avoid contact between the lens and the sample, and control over the tip/tilt angles is necessary. 

The optical element is assumed to have outer dimensions in the order of $\SI{1}{\mm}\times\SI{1}{\mm}$. The additional material outside the effective field-of-view is considered necessary for handling and manufacturing of the instrument. Combined with the piston precision, the large footprint implies a tip-tilt precision of $\le\SI{1}{\micro\radian}$.

The combination of high speed scanning and a large field-of-view is a necessary but not sufficient condition to satisfy the industry need for high throughput. The industry has indicated that there exists a current need to image $9\ \mathrm{sites/die}$ and scan $9\ \mathrm{dies/wafer}$. These numbers are, however, steadily increasing and it is expected that eventually every die on every wafer has to be inspected. To meet these metrology and inspection needs in the future, parallelization of many meta-instruments will be necessary. To be ready for future parallelization, the outer dimensions of the instrument have to be minimized. For the existing need of imaging $9$ dies on a \SI{300}{\mm} wafer, however, it suffices when the outer dimensions are limited to approximately  $\SI{70}{\mm}\times\SI{70}{\mm}$. 

\begin{table}[t]
	\centering
	\caption{Summary of system requirements.}
	\label{tab:requirements}
	\begin{tabular}{@{}lll@{}}
		\hline \hline
		\textbf{Specification} & \textbf{Threshold} & \textbf{Goal} \\ \hline
		Scan speed &  \SI{1}{\mm\per\second} & \SI{100}{\mm\per\second} \\
		Bandwidth & \SI{100}{\kHz} & \SI{10}{\MHz} \\
		Piston range & \SI{1}{\mm} & \SI{10}{\mm} \\
		Piston precision & \SI{1}{\nm} & \SI{0.1}{\nm} \\
		Tilt range & \SI{1}{\milli\radian} & - \\
		Tilt precision & \SI{1}{\micro\radian} & \SI{0.1}{\micro\radian} \\
		Dimensions & $\SI{70}{\mm}\times\SI{70}{\mm}$ & - \\
		\hline\hline
	\end{tabular}
\end{table}

\subsection{Functional layout}
\label{sec:funclayout}
\begin{figure*}
	\centering
	\includegraphics[width=15cm]{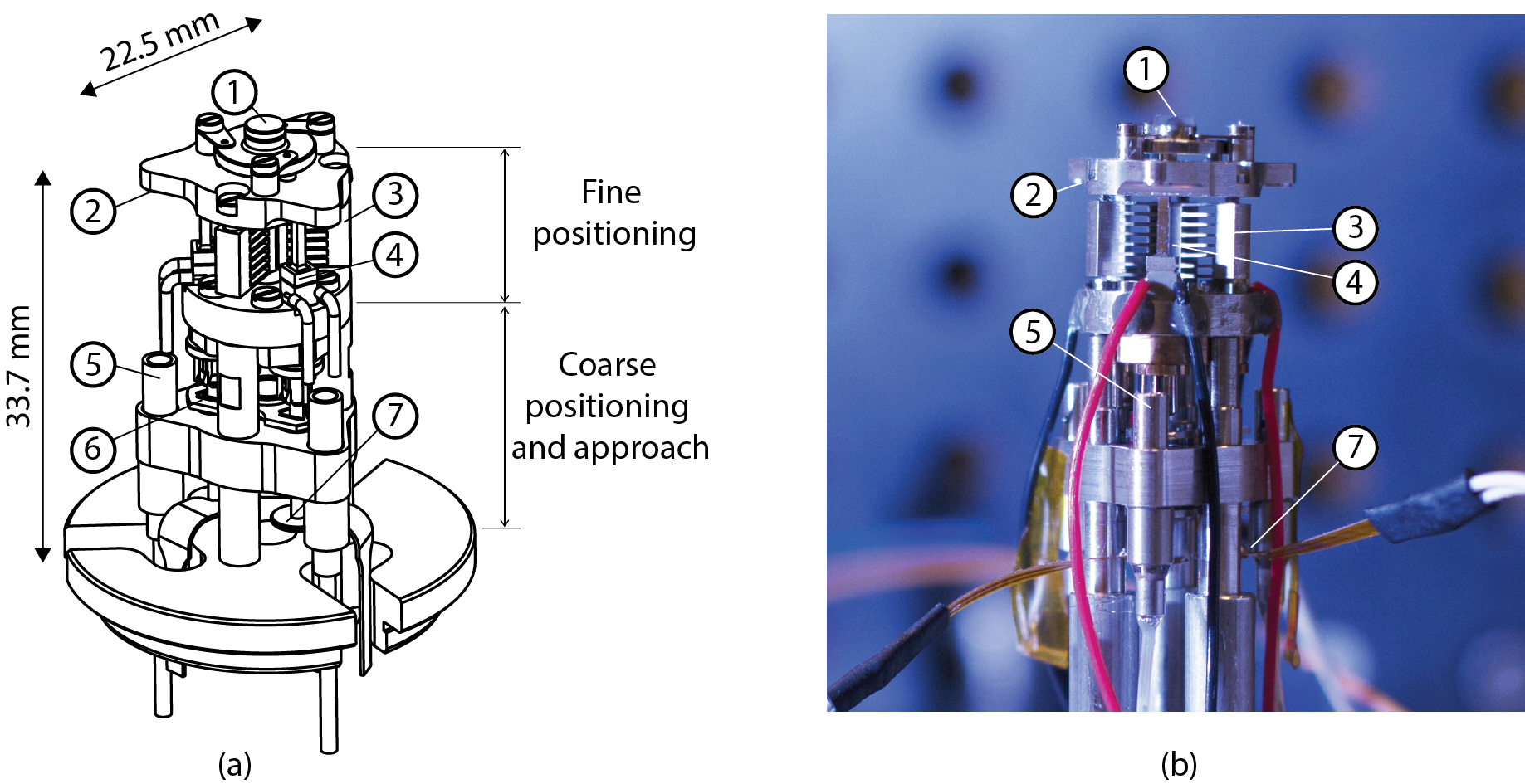}
	\caption{\label{fig:architecture}(a) Isometric drawing of the first generation meta-instrument. (b) Photo of the realized instrument. Indicated are (1) the optical element or MEMS positioning stage; (2) the mirror surfaces for interferometry; (3) the leaf springs for pre-stressing the piezo actuators in the fine positioning stage; (4) the piezo elements with struts; (5) the fiber interferometer heads; (6) the leaf springs for clamping and alignment of the coarse linear actuators (not visible in the photo) and (7) the coarse linear actuators. }%
\end{figure*}

The optical element has to be positioned at the correct distance to the sample and has to be parallel to its surface. The required degrees of freedom in piston motion and tip-/tilt- rotations are realized by a combination of two stages: a coarse positioning stage for approaching the sample and tracking coarse height changes, and a fine positioning stage for high speed tracking of small height variations. These layers are depicted in \figref{fig:architecture}. Each of the two stages is composed of three actuators that are placed at \SI{120}{\degree} angles apart to provide the tip- and tilt- degrees of freedom. 

To obtain a very high bandwidth piston motion, the favorable scaling of device mass and structural stiffness found in micro-electromechanical system (MEMS) is used to our advantage. Hyperlenses, nano-antennas and SOLs are manufactured using the same techniques that also used for creating MEMS devices, which allows for the integration of the optical element directly into the moving device to realize bandwidths between \SI{100}{\kHz} and \SI{1}{\MHz}. 

The motion of the fine positioning platform is measured by three interferometers, which are placed adjacent to the actuators at \SI{120}{\degree} angles, while the motion of the MEMS stage is measured by a capacitive sensor as explained in Section~\ref{sec:MEMS}. The distance between the lens and sample cannot be measured at the moment. This is a current challenge as further explained in Section~\ref{sec:controlGap}.

To control the position and rotation of the optical element with respect to the sample surface, the multiple inputs from the available sensors are used to drive the combination of the multiple actuators simultaneously. This multiple input, multiple output (MIMO) problem is controlled using the so-called 'offloading' strategy\cite{Witvoet2015}. Herein two fast control loops for the MEMS stage and piezo actuators, respectively, are nested in a slow control loop that controls the approach stage. Each loop cancels the DC-offset of the next faster loop. This allows for the combination of a large stroke and high precision, in combination with a high control bandwidth. 

\section{Realization}
\label{sec:current}
To satisfy the requirements that are stipulated in Section~\ref{sec:architecture}, a first iteration of the instrument was realized (see \figref{fig:architecture}) and tested. It consists of a stack of two stages to achieve the required stroke, tip/tilt range, tracking precision and bandwidth. The first layer (bottom of the image) comprises the coarse approach stage, while the second layer comprises the fine positioning stage. Both are described in more detail in Section~\ref{sec:approachfinepos}, followed by a discussion on the realized fiber interferometers in Section~\ref{sec:interferometer}. The third, (optional) MEMS stage is discussed separately in Section~\ref{sec:MEMS}. All mentioned components are highlighted and numbered in \figref{fig:architecture}.

\subsection{Approach and fine positioning}
\label{sec:approachfinepos}
The approach stage serves the purpose of bringing the instrument in close proximity to the sample surface by providing a large stroke of a few millimeters and a coarse, micrometer step size. In this way, travel speeds of $\SI{1.5}{\mm\per\s}$ are obtained that limit the time of engaging the instrument to the sample to a few seconds. 

This motion is achieved by a set of three linear actuators (TULA by PiezoTech, 485 Bonghwasan-ro, Sangbong-dong, Jungnang-gu, Seoul, South Korea), which are clamped by collets and suspended by means of three leaf-springs. The actuators consist of a piezo-electric hat, which is connected to a light weight shaft. The piezo-electric element is actuated to deliver an impulse to the shaft, which in turn moves through the collet by means of a stick-slip mechanism. The step size and step speed are controlled by tuning the contact pressure that is provided by the collets. 

The fine positioning stage uses three stacked piezo-electric actuators (Noliac, Hejreskovvej 18B, DK-3490 Kvistgaard, Denmark), that are connected to a platform by means of struts. Each piezo actuator is prestressed by means of an array of 8 leaf springs to increase the (large signal) bandwidth. The struts are connected to a platform that features an aperture on top of which either a lens can me mounted directly, or on which the nanopositioning MEMS device is placed. The struts provide sufficient compliance, such that tip- and tilt motions are possible. The three corners of the platform are polished to provide mirror targets for the fiber interferometers. 

It was found experimentally that the localized high clamping force introduced by the collets, yield unpredictable behaviour of the TULA linear actuators. This problem was solved in the next generation, by distributing the clamping force over a larger area, to reduce the contact pressure, as is discussed in more detail in Section~\ref{sec:secondgenmechanics}.

\begin{figure*}
  	\centering
  	\includegraphics[width=17cm]{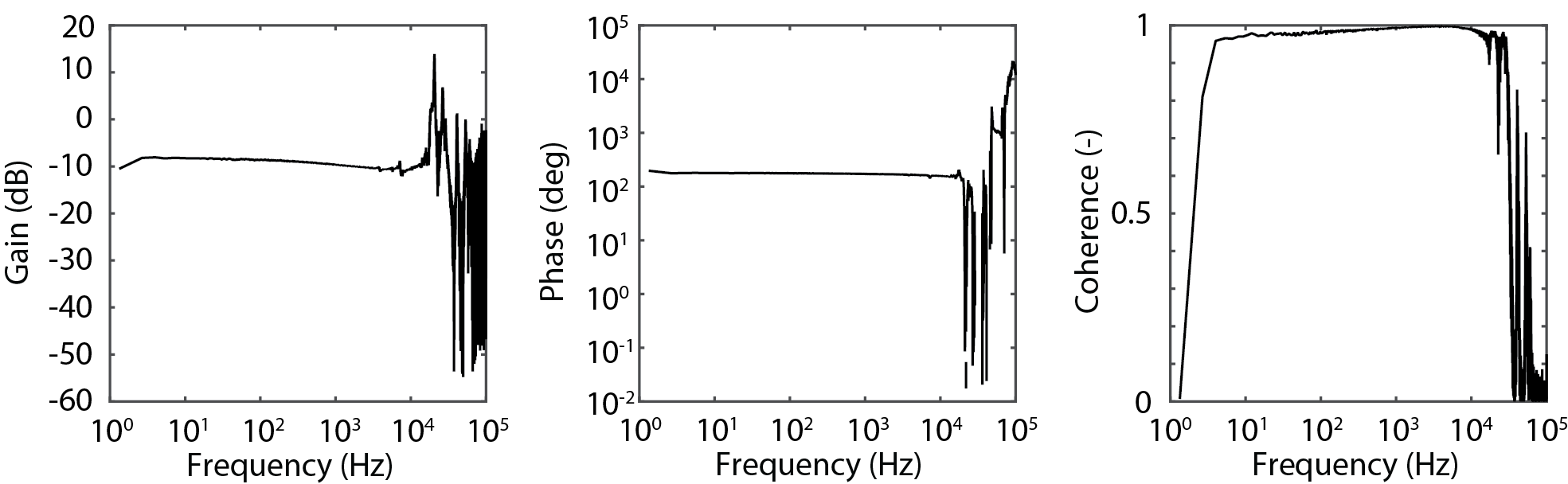}
  	\caption{\label{fig:MIFRF}Transfer function of the fine positioning stage. Shown are the gain (left), phase (center) and signal coherence (right). The transfer function shows good coherence up to \SI{10}{\kHz}. }%
  \end{figure*}

Measurements of the plant transfer function for the fine positioning stage show a \SI{30}{\dB} decoupling between the piston and the tip/tilt motions. As can be seen from \figref{fig:MIFRF}, the signal coherence remains close to 1 up to a frequency of \SI{10}{\kHz}, which in this case can be considered the theoretical limit for the control bandwidth. Slow signal drift in the interferometer signals limits the performance at lower frequencies (see Section~\ref{sec:interferometer}). It should be noted here that the coarse positioning stage was excluded from the measurement.

\subsection{Fiber interferometer}
\label{sec:interferometer}
\begin{figure}[t]
 	\centering
 	\includegraphics[width=6.5cm]{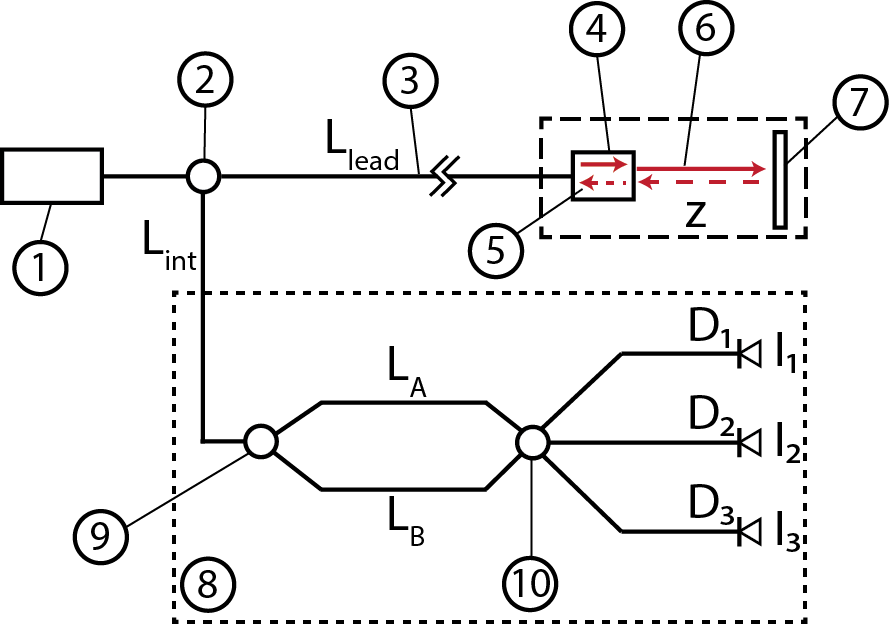}
 	\caption{\label{fig:interferometer}Schematic of the fiber interferometer layout designed and developed by TNO. Indicated are (1) the light source, (2) a beam splitter, (3) the lead fiber, (4) the sensor head, (5) the reference beam, (6) the sensing beam, (7) the target, (8) the detection interferometer unit, (9) a beam splitter and (10) a $3 \times 3$ beam combiner.}%
 \end{figure}

For the current development, a fiber optic based interferometric technology is identified to be most suitable as displacement sensor, for its small form factor and high sensitivity. The conventional optical interferometer with bulk optics is a well-known technology for high accuracy displacement sensing and is widely used in high-precision equipment, e.g., lithography machines\cite{Meskers2014}. In comparison to other type of optical displacement sensing technologies, interferometer combines sub-nm resolution with large working range and potentially a high detection bandwidth. Two-beam interferometers\cite{Renishaw2016} are already widely used in the semiconductor industry for position and movement control. The sensor heads consist of high precision bulk optical components and have in general a volume of about \SI{100}{\cm\cubed}. A new interferometer concept based on fiber optic technology was developed and patented\cite{L.K.2013} to enable the realization of a sensor head in the order of \SI{10}{\mm\cubed} with a small diameter. This configuration is shown in \figref{fig:interferometer}. This layout is included in the realized setup in threefold to provide three independent interferometers. The components discussed in this section are indicated in the figure.

The basic setup of the interferometer consists of a light source, a beam splitting device, and a lead fiber with length $L_\mathrm{lead}$ connecting to the sensor head at the measurement location. The sensor head is equipped with an internal beam splitter to generate the reference beam and the sensing beam. The reference beam is transported directly back by the lead fiber towards the beam splitting device. A lens in the sensor head is used to provide the sensing beam with beam diameter and divergence to illuminate the target which is located at a distance $z$ from the sensor head (see \figref{fig:interferometer}). The reflection of the sensing beam is coupled into the lead fiber using the same lens. Both the beam splitter and the lens in the sensor head are based on fiber optic technology. Therefore the sensor head is small, with tens of mm in length and about \SI{1}{\mm} in diameter.  

Both the reference beam and the reflection of the sensing beam from the target are transported by the lead fiber back to the beam splitting device and redirected to the detection interferometer unit at a distance $L_\mathrm{int}$ of the beam splitting device. In the detection interferometer unit, interference is generated between the reference beam and the sensing beam to be able to measure the distance $z$ with high resolution.

Starting with a beam with field $E_0$ at the beam splitting device, optical wave propagation theory can be used to described the electrical field of the reference beam and the sensing beam at the input of the detection interferometer unit. They are denoted as $E_{\mathrm{ref}}$ and $E_{\mathrm{sens}}$. The phase difference between $E_{\mathrm{ref}}$ and $E_{\mathrm{sens}}$ is equal to $2\pi\left(2z\right)/\lambda$ where $\lambda$ is the wavelength of the light used. Measuring this phase difference using the detection interferometer unit results in the detection of the target distance $z$. 

In the detection interferometer unit, the beam splitter divides both the reference beam and the sensing beam over path A and B with respective path lengths $L_\mathrm{A}$ and $L_\mathrm{B}$. This results in the fields $E_{\mathrm{ref, A}}$ and $E_{\mathrm{sens, A}}$ in path $A$, and $E_{\mathrm{ref, B}}$ and $E_{\mathrm{sens, B}}$ in path $B$. After passing path $A$ respectively path $B$, the beams are combined by a special $3 \times 3$ beam combiner and interference occurs. The  $3 \times 3$ beam combiner has $3$ outputs and $3$ detectors ($D_1$, $D_2$ and $D_3$ in \figref{fig:interferometer}) which are used to detect the optical power of the $3$ interference signals. The special feature of the $3 \times 3$ beam combiner is the mutual phase difference of $2\pi/3$ between the $3$ interference signals:
\begin{widetext}
\begin{subequations}
	\begin{equation}
	I_1 = E_\mathrm{ref, A}^2 + E_\mathrm{ref, B}^2 + 2V\sqrt{E_\mathrm{ref, A}^2E_\mathrm{sens, B}^2}\cos\left(k\left(\left(L_\mathrm{A} - L_\mathrm{B})-2z\right)\right)\right);
	\end{equation}
	
	\begin{equation}
	I_2 = E_\mathrm{ref, A}^2 + E_\mathrm{ref, B}^2 + 2V\sqrt{E_\mathrm{ref, A}^2E_\mathrm{sens, B}^2}\cos\left(k\left(\left(L_\mathrm{A} - L_\mathrm{B})-2z\right)\right) + \frac{2\pi}{3}\right);
	\end{equation}
	
	\begin{equation}
	I_3 = E_\mathrm{ref, A}^2 + E_\mathrm{ref, B}^2 + 2V\sqrt{E_\mathrm{ref, A}^2E_\mathrm{sens, B}^2}\cos\left(k\left(\left(L_\mathrm{A} - L_\mathrm{B})-2z\right)\right) - \frac{2\pi}{3}\right),
	\end{equation}
\end{subequations}

\end{widetext}
where $V$ is the visibility of the interference signal and $k$ is the wavenumber. 

The phase $\phi$ of the interference signal in a $3 \times 3$ interferometer system can be calculated as\cite{Cheng2015}:
\begin{equation}
\label{eq:phi}
\begin{split}
\phi &= k\left(\left(L_\mathrm{A}-L_\mathrm{B}\right)-2z\right)\\ 
	 &= \frac{2\pi}{\lambda}\left(\left(L_\mathrm{A}-L_\mathrm{B}\right)-2z\right)\\ 
	 &= \arctan\left(\sqrt{3}\frac{I_1-I_2}{2I_3 - I_1 - I_2}\right)
\end{split}
\end{equation}
Using the optical power of the 3 interference signals, the change in the phase of the interference signal can be calculated and the change in target distance $z$ can be determined using \eqref{eq:phi} as $z = 1/2 \left( \left(L_\mathrm{A} - L_\mathrm{B}\right) - \phi/k \right)$.

\begin{figure*}
  	\centering
  	\includegraphics[width=11.5cm]{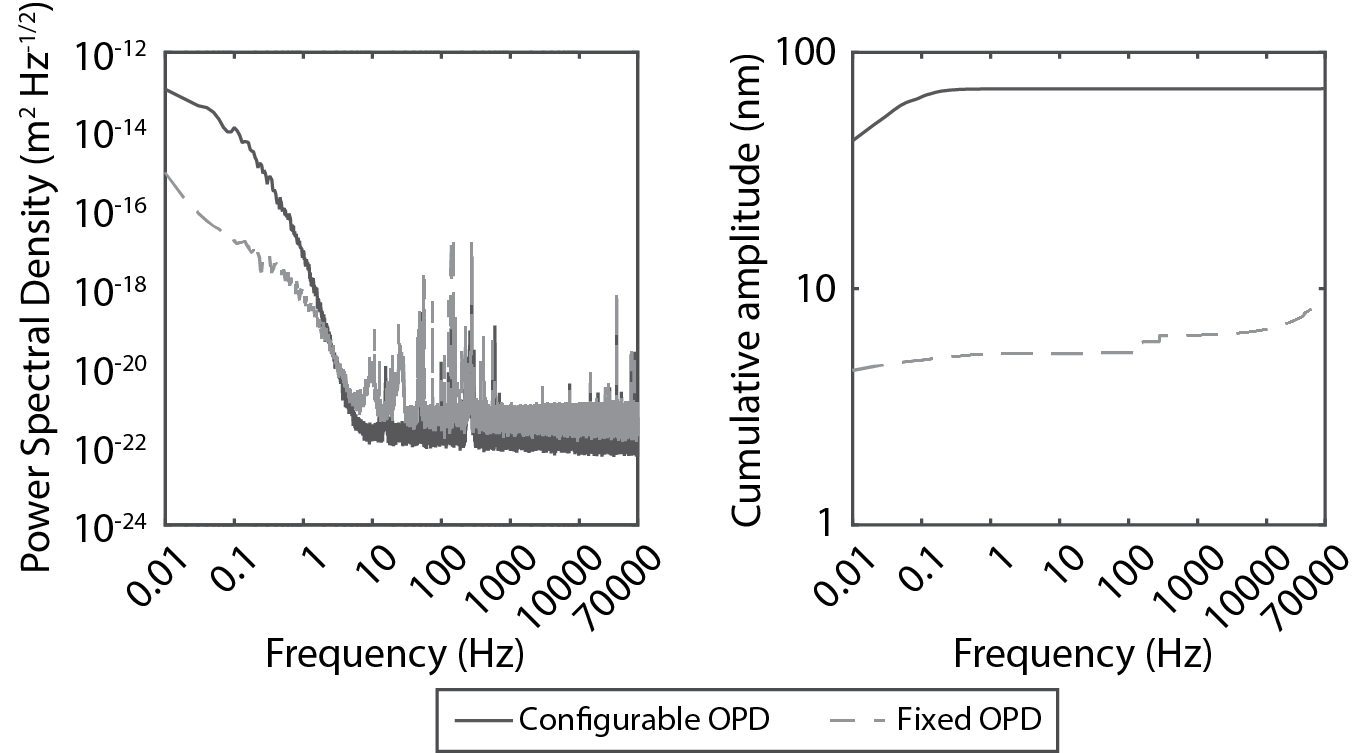}
  	\caption{\label{fig:IFMnoise}Power spectral density of the measured interferometer signals over a \SI{1000}{\second} interval (left) and the cumulative amplitude spectrum (right) for the cases in which either a configurable optical path difference (OPD) or a fixed OPD is used. It is clear from the cumulative amplitude that the bulk of the drift can be accounted to frequency components $\le\SI{1}{\Hz}$. This is typical for temperature induced effects. }
\end{figure*}

The measured noise spectral density together with the cumulative noise are shown in \figref{fig:IFMnoise}. A severe drift of \SI{70}{\nm} was measured over a \SI{1000}{\second} time span (sensor A), when an in-air interferometer was used with a configurable optical path difference (OPD). From the (inverse) cumulative amplitude spectra it is clear that the bulk of the drift is caused by components at frequencies $\le\SI{1}{\Hz}$. This is typical for temperature induced effects. When the detection interferometer is changed to an in-fiber configuration with fixed OPD and installed in an aluminum packaging to suppress impact of thermal fluctuations, the drift is reduced to  \SI{10}{\nm} for the same time span. The drift can be further reduced by using active temperature control. 


\subsection{Nanopositioning MEMS device}
\label{sec:MEMS}
High mechanical bandwidths are attainable using micro-electromechanical systems (MEMS), because of the combination of low mass and high structural rigidity, that are enabled by favorable scaling towards the micron scale. As the tip- and tilt- degrees of freedom can be controlled with sufficient resolution and accuracy by the combined coarse and fine positioning stages, the MEMS positioning stage only has to provide a piston motion. 

 
  
    \begin{figure*}
    	\centering
    	\includegraphics[width=17cm]{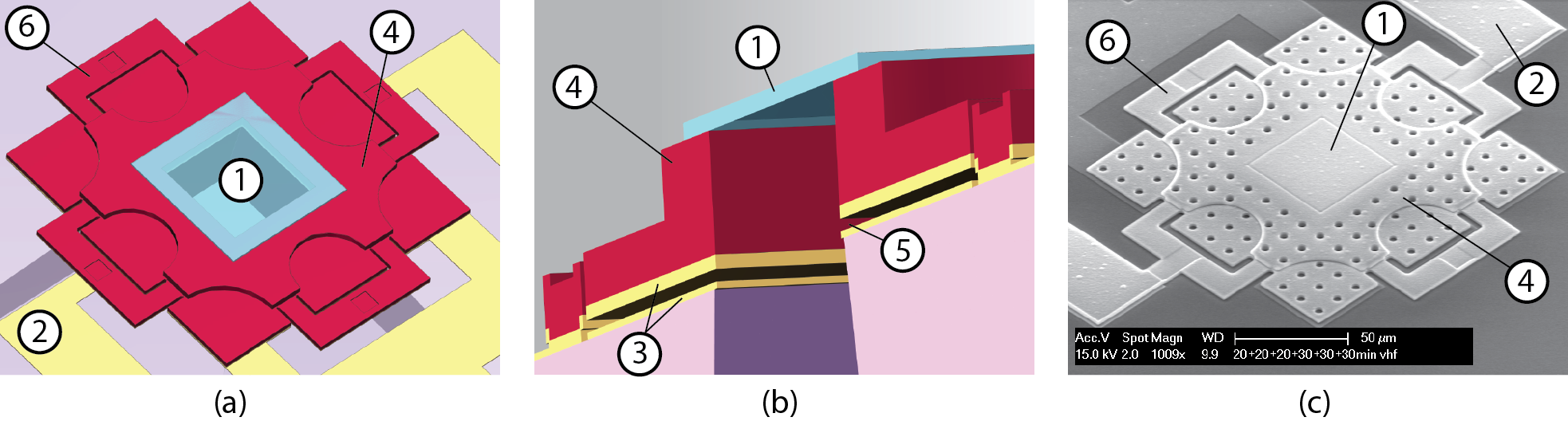}
    	\caption{\label{fig:MEMS3}(a) Isometric rendering of the micro-electromechanical system. (b) Section view. (c) scanning-electron micrograph of the manufactured device. Courtesy of Else Kooi Laboratory, TU Delft. Shown are (1) the transparant window, that can be replaced by any of the mentioned optical elements; (2) the electronic leads that are connected to (3) the electrodes; (4) the moving plate, made out of silicon carbide; (5) the air gap that separated the electrodes; and (6) the leaf springs. At the thickest section, the moving plate has a thickness of \SI{4}{\um}.}%
    \end{figure*} 

The device that we designed and realized comprises a parallel plate arrangement that consists of a fixed electrode and of an out-of-plane moving electrode, and measures $\SI{150}{\um} \times \SI{150}{\um}$ in dimensions. \figref{fig:MEMS3} a-c show an isometric top-view, section view, and a scanning electron micrograph of the MEMS stage, respectively. The out-of-plane motion is achieved by means of electrostatic actuation in which an electric potential yields an attractive force between the electrodes. The restoring force is supplied by a combination of eight L-shaped leaf-springs, which are situated at the edges of the moving electrode. The gap between the electrodes is designed to be \SI{500}{\nm}, which results in an effective stroke of \SI{166}{\nm} that is limited by the electrostatic pull-in effect\cite{Nielson2006}. The optical element is integrated into the moving plate as part of the MEMS manufacturing process and has a window of $\SI{40}{\um} \times \SI{40}{\um}$. The window is achieved by a back-etch of the structure and reached up to the optical element to allow light to pass through the stage. 

The effective bandwidth is maximized by minimizing the mass, while maximizing the structural rigidity. To this effect, the moving plate is made out of \SI{2}{\um} thick silicon carbide (SiC), which has a high specific stiffness of approximately \SI{132e6}{\meter\squared\per\second\squared}. This is an improvement of a factor \num{1.5} to \num{2} over the more conventional choices of silicon nitride or silicon, respectively. In addition to this, the plate is further stiffened by increasing the thickness near the optical element to a total of \SI{4}{\um}. 

\begin{figure}
 	\centering
 	\includegraphics[width=8.5cm]{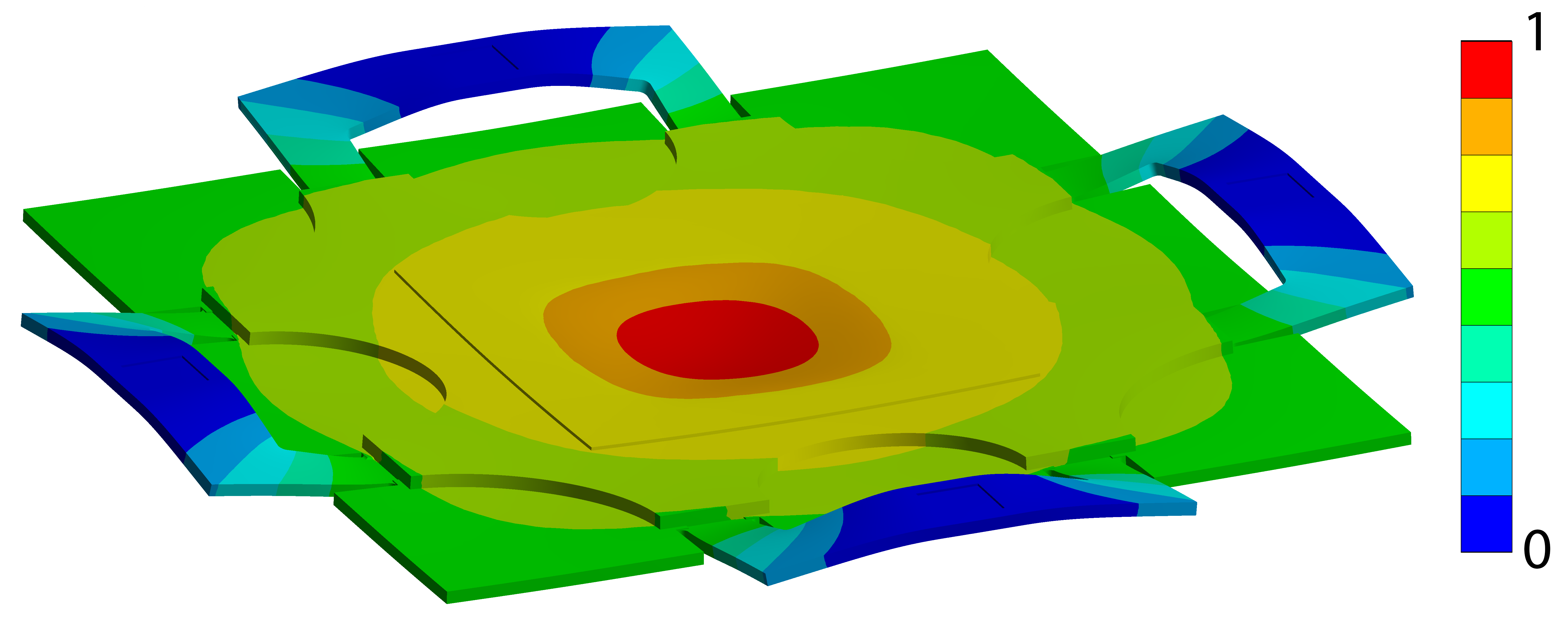}
 	\caption{\label{fig:MEMSModal1}First mode shape resulting from finite-element analysis. The mode shape has minimal flapping of the corners of the electrodes and happens at a resonance frequency of \SI{660}{\kHz}.}%
 \end{figure}

In the first bending mode of the moving electrode, its center and free corners can show a large out of plane motion. This 'flapping' motion, is minimized by moving the joint between the leaf springs and the electrode away from the extremes of the electrode and more towards its center. The exact joint locations were optimized using an iterative design procedure and checked using a combinations of modal finite-element analysis and the a static deformation analysis, in which a uniform pressure is applied to the electrode to simulate the effect of the electrostatic field. The mode shape of the first resonant mode, which occurs at \SI{660}{\kHz}, is depicted in \figref{fig:MEMSModal1}. The analysis also shows that when a constant pressure (equivalent to the distributed electrostatic force) is applied to the electrode, the displacement is almost fully achieved by deforming the leaf springs, rather than by deforming the electrode. This causes the free plate to operate primarily as a rigid body. 

To be able to control the position of the moving electrode, its location is measured by an integrated distance sensor. Although this could also be done using an optical technique, e.g. interferometry, miniaturization dictates that integrated capacitive sensing of the gap size is preferred. A bandwidth between \SI{100}{\kHz} and \SI{1}{\MHz} can be realized by means of an implementation\cite{Herfst2012} of Nieminen's approach for fast capacitive measurements\cite{Nieminen2005a}. It utilizes probe currents at a frequency of \SI{2.5}{\GHz}, which are injected via the ground-signal-ground interconnects that are present in the MEMS stage. Because these signals are far above the first mechanical resonance frequeny of \SI{660}{\kHz}, mechanical damping attenuates any mechanical response to negligible amplitudes. 

\begin{figure}
 	\centering
 	\includegraphics[width=8.5cm]{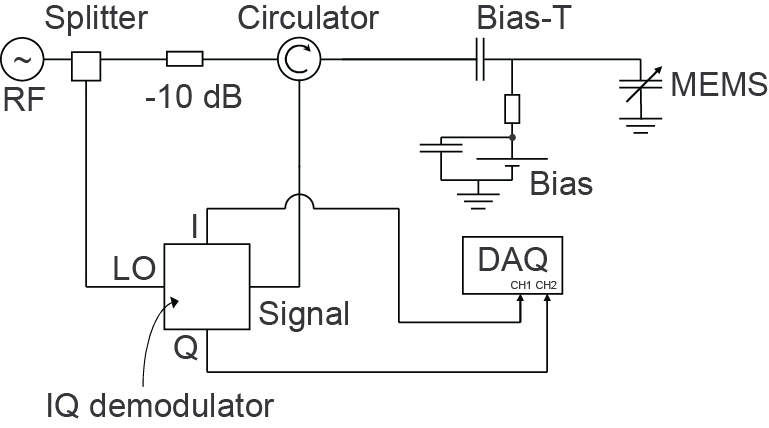}
 	\caption{\label{fig:fastrf}Schematic of the 1-port radio-frequency measurement system. Phase and amplitude of the reflected radio-frequency signal are measured with an IQ demodulator, from which the capacitance can be determined. Adapted reproduction from R.W. Herfst, P.G. Steeneken, M.P.J. Tiggelman, J. Stulemeijer, and J. Schmitz, IEEE Transactions on Semiconductor Manufacturing \textbf{25}, 310 (2012) \copyright~2012 IEEE.}%
\end{figure}

At the used gigahertz frequencies, the interconnect between the MEMS device and detection electronics acts as a waveguide. At the end of this waveguide the RF waves are reflected by the MEMS device with a phase shift that depends on the capacitance and hence the position of the moving electrode. An IQ-demodulator is used to measure the in-phase (I) and quadrature (Q) component of the reflected signal, which in turn can be used to determine the amplitude and phase. A schematic of the 1-port measurement system, is shown in \figref{fig:fastrf}. The proper matching of hardware components results in a root-mean squared position noise that is theoretically limited to $\le\SI{0.2}{\nm}$ over a bandwidth from \SI{10}{\kHz} to \SI{1}{\MHz}. In this realization, the radio frequency (RF) signal is generated by a single tune  oscillator (ZX95-2536C+ by Mini-Circuits, 13 Neptune Ave, Brooklyn, NY 11235, USA), which produces a frequency between \SI{2315}{\MHz} and \SI{2536}{\MHz}. The reflected signal is circulated by a circulator (D3C2327S by DiTom Microwave Inc., 7592 N. Maroa Ave., Fresno, CA 93711, USA) and demodulated by an IQ demodulator (QMK2450A by Synergy, 201 McLean Blvd., Paterson, NJ 07504, USA). The RF source signal is split by a  splitter (Mini-Circuits ZX10-2-42+) and we use a bias-T (Mini-Circuits ZFBT-352-FT+) to bias the actuator for actuation. 
 
\section{Second generation}
\label{sec:future}
To further increase the performance, several design changes have been implemented that resulted in a second generation of the instrument. In this section, these changes are discussed in combination with preliminary findings of their impact on the instrument performance. 

\subsection{Mechanics}
\label{sec:secondgenmechanics}
In the pursuit of a higher mechanical bandwidth, the moving mass of the instrument has to be decreased, while the stiffness has to increase. Both can be achieved by further miniaturization of the instrument.

\begin{figure}
	\centering
	\includegraphics[width=6cm]{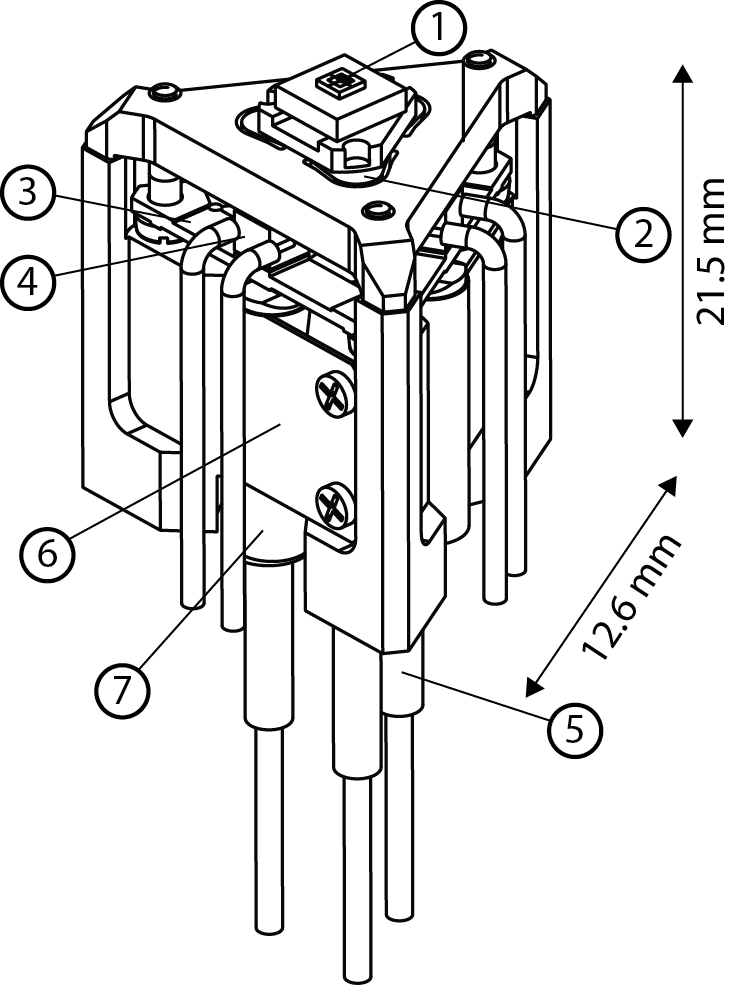}
	\caption{\label{fig:architecturenewinstrum}Isometric view of the realized meta-instrument. Indicated are (1) the optical element or the MEMS positioning stage; (2) the mirror surfaces for interferometry; (3) the leaf springs for pre-stressing of the piezo actuators in the fine positioning stage; (4) the piezo blocks; (5) the fiber interferometers; (6) leaf springs for clamping and alignment of the linear actuators in a V-groove support and (7) the carbon moving rod. These rods will be connected to the support structure. }%
\end{figure} 

By merging the functionality of the coarse approach stage and the fine positioning stage, the overall length of the instrument and the moving mass are reduced. The original coarse linear actuators are replaced by carbon tubes, which are attached to the piezo-electric actuators of the fine positioning stage. 
In the original design, the rods of the linear actuators are suspended at two points. This resulted in a relatively low first resonance mode, in which the rods were free to bend between their suspension points. By a combined action of a higher bending stiffness of the carbon rods and by clamping them in a V-groove, this mode is suppressed and the first resonance frequency is increased. The choice of a V-groove over the collets, allows for a lower contact pressure per unit length for the same contact force, which reduces wear and distributes the effect of surface roughness on the stick-slip action over a larger area for better reproducibility. 

The piezo-electric actuators are driven using a control voltage at a quadratic rise function, which results in a constant driving force. At a frequency of $\ge\SI{10}{\kHz}$, the stage operates in an 'impulsive hammer' mode.  Using a stick-slip mechanism, the moving mass translates along the rods in sub-micrometer sized steps. Below this \SI{10}{\kHz} threshold, the stage operates in the original fine positioning mode, which can be used to track the surface topography with nanometer precision.

\begin{figure*}
  	\centering
  	\includegraphics[width=16.5cm]{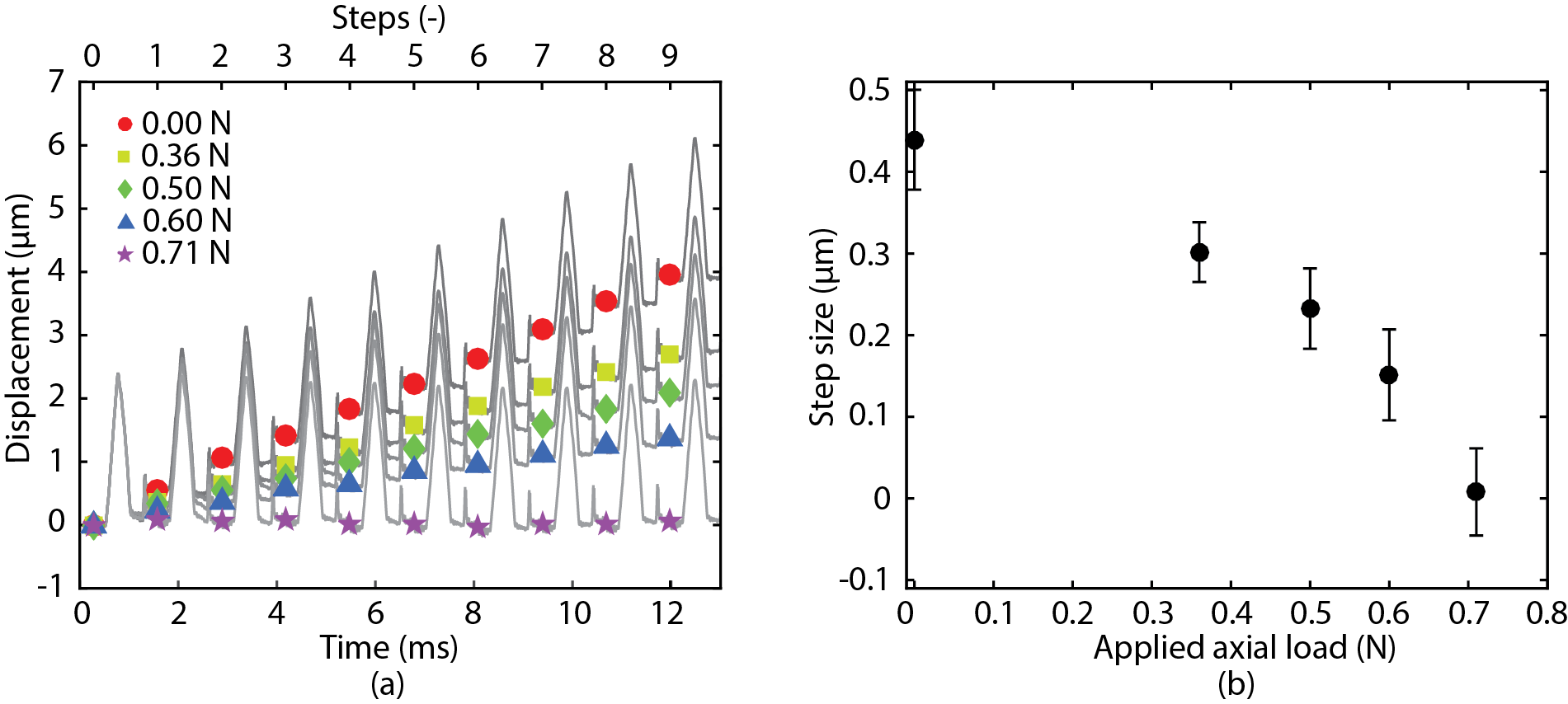}
  	\caption{\label{fig:TNOtulasteps} (a) Measured displacement as function of axial load. Piezo element was actuated at \SI{769}{\Hz} at a peak voltage of \SI{50}{\volt}. (b) Average step size as function of axial load. The error bars indicate the standard deviation of the population for the data presented in (a). }%
  \end{figure*}

Initial tests of an isolated linear actuator show that the step size is dependent on the axial blocking force applied to the shaft (\figref{fig:TNOtulasteps}). Finite-element analysis shows that the required \SI{1}{\milli\radian} rotation range can be achieved by loading two of the piezo actuators at \SI{0.27}{\mN} in opposite directions. At this blocking force, a step size of \SI{0.32}{\um} can be achieved. At \SI{10}{\kHz} this yields a speed of \SI{3.2}{\mm\per\second}. 

In contrast to the first design that uses arrays of aluminum leaf springs to prestress the piezo elements, the new design does this using a single titanium leaf spring per piezo element. In this way, the prestress on the piezo elements can be increased to yield a higher dynamic range. This results in an increase of the first resonance mode of the approach motor from \SI{1.1}{\kHz} to \SI{8}{\kHz}.

Beside increasing the bandwidth, the merging of functionality of both stages also significantly reduces the total number of parts and the amount of cabling. The nominal instrument size is further reduced from \SI{33.7}{\mm} to \SI{22.0}{\mm} in length and from \SI{220}{\mm\squared} to \SI{158}{\mm\squared} in footprint. This allows for tighter packing of the instruments for parallel operation. To put this in perspective, this allows for $450$ concurrent measurement sites on a \SI{300}{\mm} wafer. 

\subsection{Interferometry}
The passive fiber optic components in the current setup are commercial off-the-self components for the telecommunications industry and are widely available. Since the phase is determined from the measured optical power $I_1$, $I_2$ and $I_3$, the phase resolution depends on the noise of the optical power measurement. This includes the detector noise and the dynamic range of the analog-to-digital converter. 
The change in the target distance $z$ is calculated via \eqref{eq:phi}. The stability and noise of $L_\mathrm{A}$, $L_\mathrm{B}$ and $\lambda$ affect directly the calculation of $z$ from the phase $\phi$. The components in the detection interferometer unit are all fiber optic based. Therefore, path $L_\mathrm{A}$ and $L_\mathrm{B}$ are sensitive to temperature and mechanical vibration. For the realization of the second generation, proper selection and conditioning of the fiber optic components will be required to stabilize the paths $L_\mathrm{A}$ and $L_\mathrm{B}$. Furthermore, a dedicated light source will be selected and optimized to achieve a stable wavelength. 

\section{Challenges}
The development of the meta-instrument opto-mechatronic platform is the first step towards developing a near-field imaging instrument. However, several other challenges need to be addressed before a working prototype of an imaging instrument can be realized as described in the following.

\subsection{Lens-sample interactions}
Lens-sample interactions make the positioning of thin membrane optical structures at extreme proximity to a sample a non-trivial enterprise. This a relevant concern especially for structures as the ones required for flat hyperlenses and nano-antennas, but of no concern for solid-immersion lenses and superoscillatory lenses.

For example, the low bending stiffness of such structures makes them very susceptible to the strong out-of-plane loads that originate from various sources. On one hand, there are the intermolecular interactions that manifest themselves in terms of electrostatic loads and Van der Waals/Casimir forces. Batra \etal\cite{Batra2008} and Broer \etal\cite{Broer2013} have determined the mechanical conditions under which these interactions can cause the membrane to collapse towards the sample. The optics and mechanics of the membrane will have to be considered concurrently to arrive at a feasible design. The non-linear forces that are involved in these mechanical interactions can also result in other mechanical instabilities and increased damping that may limit the dynamic performance of the positioning system. This necessitates further research of the system dynamics that includes modeling of the lens-sample interactions. 

On the other hand there are interactions with the environment that manifest only in small cavities. One example of this is capillary condensation that can cause a water bridge to form between the two surfaces for separations smaller than \SI{20}{\nm}\cite{Li2006, VanZwol2008, VanZwol2009}. This too can cause the surfaces to collapse and cause damage to both. Although the extent of this issue can be limited by removing the water from both surfaces, e.g., by performing a bake-out at high temperature at ultra high vacuum (UHV), performing such cleansing is not always possible under operational conditions. 

The last remaining lens-sample interaction of concern is the electric breakdown at the (sub)micronscale. Experiments\cite{Torres1999,Ono2000,Dhariwal2000,Wallash2003,Hourdakis2006,Strong2006} and theoretical modeling\cite{Go2010,Go2014,Loveless2016} show that the often used Paschen function, that relates breakdown voltage to distance between two bodies, is not valid at distances smaller than approximately \SI{5}{\um}. At smaller distances the breakdown voltage quickly drops, rather than increases. Once electric breakdown occurs, the discharge of energy and the associated high currents can locally damage the involved surfaces by melting. Electric breakdown can only be avoided by proper grounding of both the sample and the lens. It is imperative that charging of the elements is avoided or that the optical element is retracted to a safe distance once the onset of discharge is detected to minimize the risk of damage.

To the best of our knowledge, large scale experiments of the listed phenomena have yet to be conducted with relevant geometries consisting of thin membrane structures that span tens of micrometers at nanometer gaps. Further research is necessary to gain a better understanding of the lens-sample interactions to provide design rules and better mitigation strategies.

\subsection{Control of the gap}
\label{sec:controlGap}
Maintaining the artifact at its working distance requires the combination of high speed sensors and high speed actuators. Using the embedded interferometers it is currently not possible to measure the distance between the lens surface and the sample surface. This distance can be measured by a sensing system that is installed close to the optical element or is integrated in the optical architecture. For example, live feedback of the produced image can provide the means for determining whether the optics are positioned at the right distance. Alternative solutions can rely on point-probe techniques that rely on the dynamic response of micro-cantilever beams\cite{Maturova2016} or on the heat transferred between a probe and the sample surface \cite{Bijster2016}. The microscopic probes used in such measurements can form an integral part of the MEMS stage or the optical element and can be used to measure the lens-to-sample distance close to the point-of-interest.

\section{Conclusion}
\label{sec:conclusions}
We have presented two designs towards realizing the meta-instrument optomechatronic platform. By offering high bandwidth control of nanometer sized gaps, the industrial application of novel, near-field optical techniques is one step closer. The three tier architecture of motion control enables a large control bandwidth for tracking small height variations in the surface topography and a large stroke for engaging the sample within a few samples. 
Future challenges lie in understanding the lens-sample interactions and in measuring and controlling the lens-sample separation at the point-of-interest. 


%

\begin{acknowledgments}
This program is supported by the TNO early research program 3D Nano Manufacturing Instruments. The authors thank Jia Wei, Gregory Pandraud and Henk van Zeijl of the Else Kooi Laboratory of TU Delft for their efforts in manufacturing the MEMS device. 
\end{acknowledgments}

\bibliography{manuscript}

\end{document}